\documentclass{sig-alternate-2013}
\usepackage{amsmath}
\usepackage{amssymb}
\usepackage{algorithmic}
\usepackage{graphicx}
\usepackage{url}

\permission{Copyright is held by the International World Wide Web Conference Committee (IW3C2). IW3C2 reserves the right to provide a hyperlink to the author's site if the Material is used in electronic media.}
\conferenceinfo{WWW'14 Companion,}{April 7--11, 2014, Seoul, Korea.} 
\copyrightetc{ACM \the\acmcopyr}
\crdata{978-1-4503-2745-9/14/04. \\
http://dx.doi.org/10.1145/2567948.2579357}

\begin{document}

\title{Efficient Network Generation Under General Preferential Attachment}

\numberofauthors{3}

\author{
  \alignauthor
  James Atwood\\
  \affaddr{School of Computer Science}\\
  \affaddr{University of Massachusetts Amherst}\\
  \affaddr{Amherst, MA, 01003}\\
  \email{jatwood@cs.umass.edu}
  \alignauthor
  Bruno Ribeiro\\
  \affaddr{School of Computer Science}\\
  \affaddr{Carnegie Mellon University}\\
  \affaddr{Pittsburgh, PA, 15213}\\
  \email{ribeiro@cs.cmu.edu}
  \alignauthor
  Don Towsley\\
  \affaddr{School of Computer Science}\\
  \affaddr{University of Massachusetts Amherst}\\
  \affaddr{Amherst, MA, 01003}\\
  \email{towsley@cs.umass.edu}
}



\maketitle

\begin{abstract}

Preferential attachment (PA) models of network structure are widely
used due to their explanatory power and conceptual simplicity.  PA
models are able to account for the scale-free degree distributions
observed in many real-world large networks through the remarkably
simple mechanism of sequentially introducing nodes that attach
preferentially to high-degree nodes.  The ability to efficiently
generate instances from PA models is a key asset in understanding both
the models themselves and the real networks that they represent.  Surprisingly,
little attention has been paid to the problem of efficient instance
generation.  In this paper, we show that the complexity of generating
network instances from a PA model depends on the preference function
of the model, provide efficient data structures that work under any
preference function, and present empirical results from an
implementation based on these data structures. We demonstrate that, by
indexing growing networks with a simple augmented heap, we
can implement a network generator which scales many orders of
magnitude beyond existing capabilities ($10^6$ -- $10^8$ nodes).  We
show the utility of an efficient and general PA network generator by
investigating the consequences of varying the preference functions of
an existing model.  We also provide ``quicknet'', a freely-available
open-source implementation of the methods described in this work.

\end{abstract}

\section{Introduction}
\label{sec:introduction}
There is a clear need for scalable network generators, as the ability
to efficiently generate instances from models of network structure is
central to understanding both the models and the real networks that
they represent.  Ideally, researchers of communication and social networks
should be able to generate networks on the same scale as the real
networks they study, and many interesting networks, such as the World
Wide Web and Facebook, have millions to billions of nodes.
Furthermore, network generation is the primary tool both for
empirically validating the theoretical behavior of models of network
structure and for investigating behaviors that are not captured by
theoretical results.  The generation of very large networks is of
particular importance for these tasks because theoretically derived
behavior is often asymptotic.


However, the generation of large networks is difficult because of its high complexity.  In the case of preferential attachment (PA), arguably the most widely used generative model of networks, a non-local distribution over node degrees must be both sampled from and updated at each time-step.  If we naively index this distribution, we will need to update every node at every time-step, which implies that generating a network will have complexity of at least $O(|V|^2)$.


PA models are of particular interest because they account for the scale-free distribution of degree observed in many large networks \cite{Newman:2003da}.  For instance, scale-free degree distributions have been observed in the World Wide Web \cite{Barabasi:1999kf,Barabasi:2000ce,Broder:2000tf}, the Internet \cite{Chen:2002ta,Faloutsos:1999dp,Vazquez:2002bo}, and telephone call graphs \cite{Aiello:2000kp,Aiello:2001wt}, bibliographic networks \cite{deSollaPrice:1965vs} and social networks \cite{Ribeiro:2010cr}.


Preferential attachment models generate networks by sequentially introducing nodes that prefer to attach to nodes with high degree. While many extensions to this model class exist, all members share the same basic form: At each time-step, sample a node from the network with probability proportional to its degree; introduce a new node to the network; and add an edge from the new node to the sampled node.  This behavior has important implications for implementation.  First, PA is inherently sequential, because the next action taken depends on the state of the network, and the state of the network changes at each time-step. This implies that the algorithm is not easily parallelized.  Second, network nodes must be indexed such that they can be efficiently sampled by degree, and, because we are introducing a new node at each time-step, the index must also support efficient insertion.  Third, the relevant distribution over nodes is non-local, in that the introduction of a new node and edge affects the probability of every node in the network through the normalization factor.

Much of the work in modeling network structure has focused on the asymptotic regime.  A model is defined, and a limiting degree distribution (as $|V|$ approaches infinity) is obtained analytically.  Less effort has been focused on generating finite networks.  In the following sections, we provide a robust framework for generating networks via PA.  This framework easily scales to millions of nodes on commodity hardware.  We also provide ``quicknet'', a freely-available open-source C implementation of the framework\footnote{\url{https://github.com/hackscience/quicknet}}.




The remainder of the paper is structured as follows.  In Section \ref{sec:complexity}, we present an analysis of the complexity of generating networks from PA models.  Section \ref{sec:implementation} describes candidate methods for efficiently implementing preferential attachment generators and presents results from a simulator which implements them.  Section \ref{sec:applications} describes several applications of a PA network generator which scales to many millions of nodes.  We describe related work in Section \ref{sec:relatedwork} and present conclusions and future work in Sections \ref{sec:conclusion} and \ref{sec:futurework}, respectively.

\section{Complexity}
\label{sec:complexity}
In this section, we describe the two existing PA models as examples.  We then provide a more formal definition of a PA model, which is followed by an analysis of the complexity of generating networks from PA models.
\subsection{Description of Considered Models}
\subsubsection{Price's Model}
\begin{figure}
  Price($n$,$\lambda$):
  \small\begin{algorithmic}
    \STATE $G$ $\gets$ $G_0$
    \FOR{$i$ $\gets$ 1 to $n-|G_0|$}
    \STATE existing\_node $\gets$ sample\_in\_degree($G$, $\lambda$)
    \STATE new\_node = add\_node($G$)
    \STATE add\_edge($G$, new\_node, existing\_node)
    \ENDFOR
    \RETURN $G$
  \end{algorithmic}
  \vspace*{8pt}
  \normalsize
  Krapivsky($n$, $p$, $\lambda$, $\mu$):
  \small\begin{algorithmic}
    \STATE $G$ $\gets$ $G_0$
    \WHILE{$|V| < n$}
    \STATE $u$ $\gets$ uniform draw
    \IF{$u<p$}
    \STATE existing\_node $\gets$ sample\_in\_degree($G$, $\lambda$)
    \STATE new\_node = add\_node($G$)
    \STATE add\_edge($G$, new\_node, existing\_node)
    \ELSE
    \STATE existing\_node\_tail $\gets$ sample\_in\_degree($G$, $\lambda$)
    \STATE existing\_node\_head $\gets$ sample\_out\_degree($G$, $\mu$)
    \STATE add\_edge($G$, existing\_node\_tail, existing\_node\_head)
    \ENDIF
    \ENDWHILE
    \RETURN $G$
  \end{algorithmic}
  \caption{Generating a network with $n$ nodes under Price and Krapivsky's models.  $G_0$ is some small seed network.  $\lambda$ and $\mu$ are scalers which give the fitness of nodes for incoming and outgoing edges, respectively.}
  \vspace*{-10pt}
  \label{bothalg}
\end{figure}
Figure \ref{bothalg} describes Price's algorithm.  Briefly, at each time-step, a node is sampled from the network with probability proportional to its in-degree; a new node is introduced to the network; and a directed edge is added from the new node to the sampled node.  Notice that a node is added at each time-step, so that the generation of a network with $|V|$ nodes takes $|V|$ steps.
\subsubsection{Krapivsky's Model}
%
Figure \ref{bothalg} also describes the algorithm of Krapivsky et al.  At each step, the algorithm of Price's model is followed with probability $p$, and a ``preferential edge step'' is taken with probability $1-p$.  During a preferential edge step two nodes, $n_o$ and $n_i$, are sampled from the network by out- and in-degree, respectively, and an edge is added from $n_o$ to $n_i$.  Note that a node is no longer added at every step; rather, a node is added at a given step with probability $p$.  This implies that the number of iterations required to generate a network with $|V|$ nodes is a random variable with expected value $|V| / p$.  $|V| / p$ is $\Theta(|V|)$ $\forall p$, so asymptotically this is no different than Price's model.  More generally, the number of iterations 
required to generate a network with $|V|$ nodes via a PA model is $\Theta(|V|)$.

\subsection{Definitions}
In this section we provide a framework for representing general
preferential attachment models.  Note that the idea of a general PA
model is not new to this work, and that the formulation presented here
is only used to facilitate algorithmic analysis.  For a detailed
treatment of general preferential attachment, please see `The
Organization of Random Growing Networks' by Krapivsky and Redner
\cite{Krapivsky:2001bg}.



Let $G_t = (V_t,E_t)$ be the network that results from $t$ iterations of a PA simulation.  $V_t$ is the set vertices (or nodes) within the network and $E_t$ is the set of edges between elements of $V_t$.  Let $T(G_t)$ be the worst-case time complexity of generating $G_t$; that is, the worst-case time complexity of a preferential attachment simulation of $t$ iterations. 

Recall that the number of iterations required to generate a network with $|V|$ nodes via PA is $\Theta(|V|)$.  Accordingly, we will omit $t$ and frame our discussion of complexity $T(G)$ in terms of $|V|$.  


Let $A = \{a_1, a_2, ..., a_{|A|}\}$ be a set of attributes that can be defined on a network node. Let $X_v = \{x_{va_1}, x_{va_2}, ..., x_{va_{|A|}}\}$\\$\in \mathbb{R}^{|A|}$ be a setting of $A$ for node $v \in V$, and let $\lambda_{va_i} \in \mathbb{R}$ be the fitness of node $v$ for attribute $a_i$. Let
\begin{align*}
f = \left\{f_{a_{i}}(x_{va_i},\lambda_{va_i}): \mathbb{R} \times \mathbb{R} \rightarrow \mathbb{R}^+ \:|\: a_i \in A\right\}
\end{align*}
be a set of functions, where $f_{a_i} \in f$ maps $x_{va_i} \in \mathbb{R}$ and $\lambda_{va_i} \in \mathbb{R}$ to a preference mass $\mu_{va_i} \in \mathbb{R}^+$.  The ``preference mass'' $\mu_{va_i}$ is a non-negative real value that is proportional to the probability of selecting $v$ by $a_i$ under the PA model.  We will refer to the elements of $f$ as the ``preference functions'' of the PA model.


A PA model has one or more preference functions.  Price's model, for example, has a single linear preference function.  Krapivsky's model has two: one for in-degree and another for out-degree.  A ``linear preferential attachment model'' only admits linear preference functions of the form $g(x,\lambda) = c_1 x + \lambda$, a ``quadratic preferential attachment model'' only admits quadratic preference functions of the form $g(x,\lambda) = c_2 x^2 + c_1 x + \lambda$, and so on.

\subsection{Derivation of Generation Complexity}
We obtain a trivial lower bound on $T(G)$ by noting that, in order to generate $G$, we must at the very least output $|V|$ nodes, so $T(G) = \Omega(|V|)$.

A discussion of the upper bound follows.  Recall that the salient
problem in generating networks from a PA model is
indexing the network's nodes in such a way that sampling, insertion,
and incrementation can be accomplished efficiently.  Tonelli et
al. \cite{Tonelli:2010uv} provide a clever method for
accomplishing all three tasks in constant time, provided that the
preference function is linear and the fitness is both uniform across
all nodes and constant.  Given constant insertion and sampling, the generation of a
network with $|V|$ nodes takes O($|V|$) time.  Considering that the
lower bound is $\Omega(|V|)$, we have the asymptotically tight bound
of $T(G) = \Theta(|V|)$.

However, this method does not extend to
nonlinear preferential attachment.  We can improve performance by
shifting to data structures which provide $O(\text{log} |V|)$
insertion, sampling, and incrementation, giving an overall complexity
of $T(G) = O(|V| \text{log} |V|)$.

\begin{figure}
  \centering
  \includegraphics[scale=0.2]{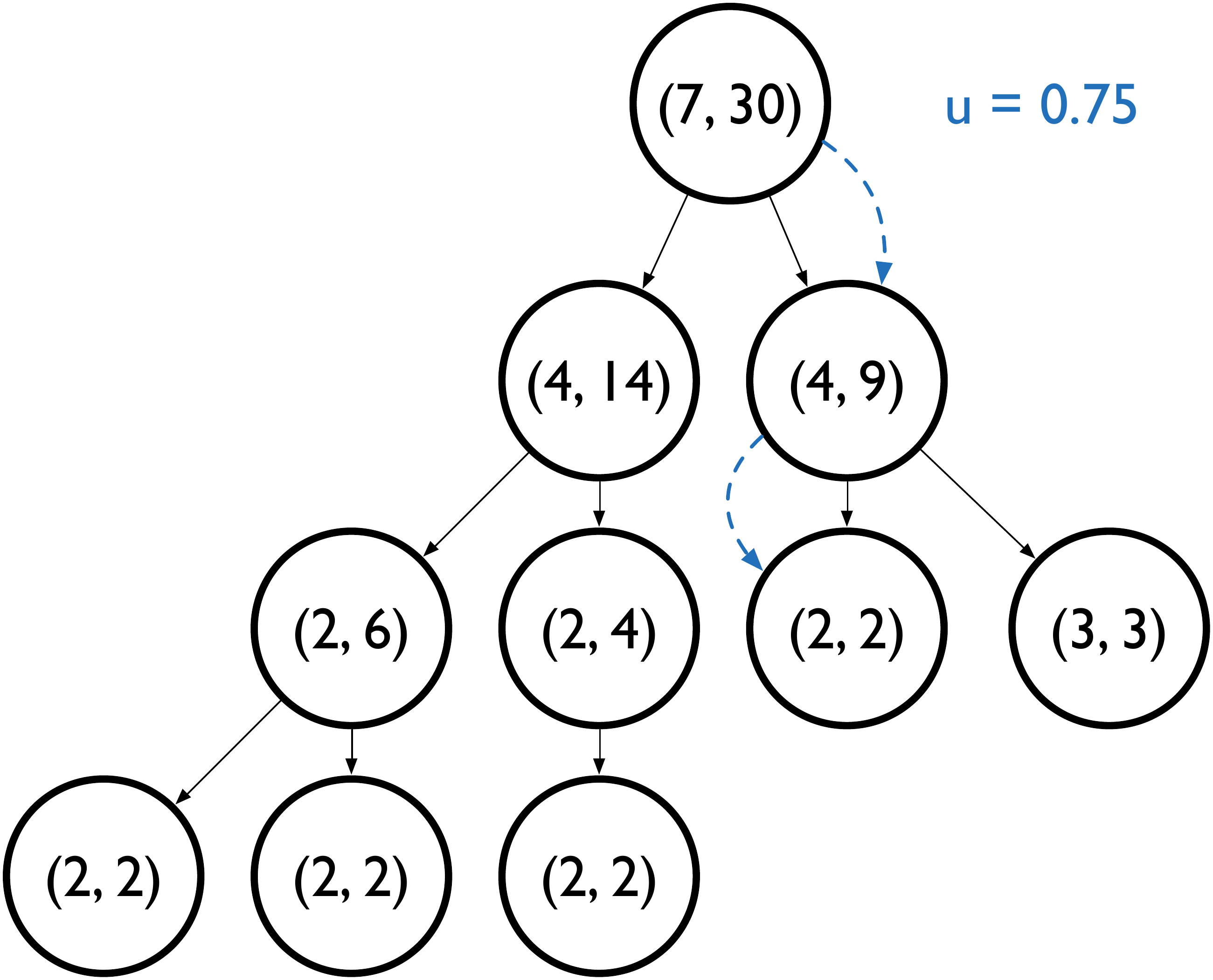}
  \caption{An example of an augmented tree structure.  Each node is annotated with $(\mu_n, \mu_s)$, where $\mu_n$ the node mass and $\mu_s$ is the subtree mass.  The preference function associated with this tree is $f(d) = d + 2.0$.  The sample path through the tree structure is illustrated for $u=0.75$.}
  \label{treeindexer}
\end{figure}

We accomplish this with a set of augmented tree structures.  Each tree supports a preference function of the model by indexing the preference mass assigned to each node in the network by that preference function.  Each item in the tree indexes a node in the network.  The tree items are annotated with the preference mass of the network node under the preference function, and the subtree mass, which is the total preference mass of the subtree that has the item as root; see Figure \ref{treeindexer}.  Note that we refer to ``items'' in the tree rather than the more typical ``nodes''; this is to avoid confusion between elements of the tree and elements of the network. We can sample from such a structure by recursively comparing the properly normalized subtree mass of a given item and its children to a uniform random draw; see Figures \ref{generalsamplealgorithm} and \ref{treeindexer}.


Note that, at each iteration of a standard PA simulation, we must sample a node, update that node's mass, and insert a new node.  In what follows we show that each of these steps can be accomplished in asymptotically logarithmic time.


\begin{figure}
  Sample(tree):
  \small\begin{algorithmic}
    \STATE sampled\_node $\gets$ NULL
    \STATE $u \gets$ uniform sample
    \IF{tree.root != NULL}
    \STATE sampled\_node $\gets$ SampleItem(tree, tree.root, 0., $u$)
    \ENDIF
    \RETURN sampled\_node
  \end{algorithmic}
  \vspace*{8pt}
  \normalsize
  SampleItem(item, $\eta$, $u$):
  \small\begin{algorithmic}
    \IF{item.left != NULL}
    \IF{$u <$ ($\eta$ $+$ item.left.subtree\_mass) / tree.total\_mass}
    \RETURN SampleItem(tree, item.left, $\eta$, $u$)
    \ENDIF
    \STATE $\eta$ $\gets$ $\eta$ $+$ item.left.subtree\_mass
    \ENDIF
    \STATE $\eta$ $\gets$ $\eta$ $+$  item.node\_mass
    \IF{$u$ $<$ observed\_mass / tree.total\_mass}
    \RETURN item.node
    \ENDIF
    \IF{item.right != NULL}
    \RETURN SampleItem(tree, item.right, $\eta$, $u$)
    \ENDIF
  \end{algorithmic}
  \caption{General algorithm to sample from the augmented tree structure.  $\eta$ is the mass observed thus far and $u$ is a sample from the standard uniform distribution.}
  \vspace{-10pt}
  \label{generalsamplealgorithm}
\end{figure}


\section{Implementation}
\label{sec:implementation}

The tree structure that we described in the previous section can be implemented in a number of different ways that have a generation time of $O(|V| \text{log} |V|)$.  They differ in their computational time for finite $|V|$.  In this section, we empirically evaluate a set of realizations of the annotated tree structure. Specifically, we investigate a simple binary max-heap where priority is defined by node mass, and a set of binary treaps with various sort and priority keys.

Note that, in the discussion of the heap-based and treap-based implementations of the tree structure, we will often refer to a ``sort invariant'' and a ``heap invariant''.  The sort invariant states that, for any three nodes $Y \leftarrow X \rightarrow Z$ where $Y$ and $Z$ are the left and right children of parent $X$, respectively, and a ``sort key'' $k$ that is associated with each item, $Y.k \leq X.k \leq Z.k$.  The heap invariant states that for any three nodes $Y \leftarrow X \rightarrow Z$ (defined in the same fashion) and some ``priority key'' $p$ associated with each item, $X.p \geq Y.p$ and $X.p \geq Z.p$.

\begin{figure}
  \centering
  \includegraphics[scale=0.2]{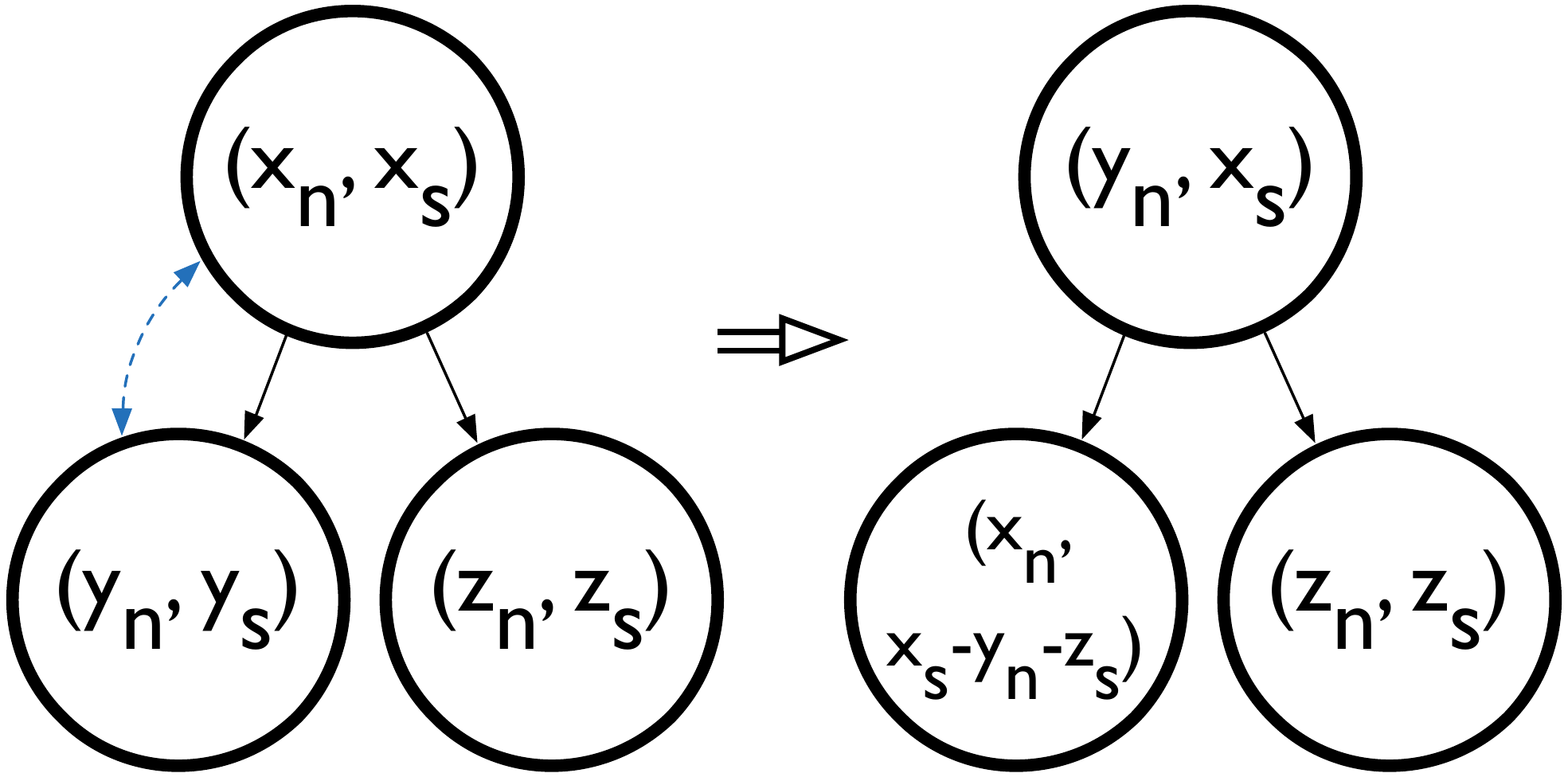}
  \caption{A diagram of the heap exchange process in the augmented heap.  Each node is annotated with $(\mu_n, \mu_s)$, where $\mu_n$ the node mass and $\mu_s$ is the subtree mass.  Note that exchanges maintain the subtree mass invariant.}
  \vspace*{-10pt}
  \label{heapexchange}
\end{figure}


We first describe the binary maximum heap.  We annotate each item in the heap with a node mass, which is defined by the preference function, and a subtree mass, which is initialized to the node mass.  When inserting a new item $i$, $i$'s node mass is added to all traversed items, so that the subtree mass is remains accurate upon insertion. Sampling is accomplished via the algorithm of Figure \ref{generalsamplealgorithm}.  Node mass may only increase, so we implement an augmented version of increase-key which maintains subtree mass under exchanges; see Figure \ref{heapexchange} for a diagram of the exchange operation.  The increase-key operation supports the Increment operation, which is described below.  We set priorities to be equivalent to node masses so that the most probable nodes can be accessed more quickly.

The PA process is supported by the binary maximum heap via the operations Sample, Increment and Insert.  As previously mentioned, sampling is performed via the algorithm of Figure \ref{generalsamplealgorithm}.  Increment increases an item's mass and then performs heap exchanges to account for any violation of the heap invariant; it is described in Figure \ref{heapincrement}.  Insert adds an item to the index, appropriately updating the subtree masses of any parent items; see Figure \ref{heapinsert}.

Note that, if we were to annotate each item with a node's \emph{probability} mass rather than \emph{preference} mass, insertion would be a linear time operation.  When a new node is introduced, the probability of every existing node decreases because the normalization factor increases.  Thus, upon insertion, every item's probability mass would need to be updated.  There are $|V|$ items, so insertion becomes a $\Theta(|V|)$ operation in this situation.  Conversely, the preference mass of each node is unaffected by the introduction of a new node.  Insertion in this scenario is a $O(\text{log}|V|)$ operation; see Figure \ref{heapinsert}.

\begin{figure}
  Increment(heap,item,new\_mass):
  \vspace*{-8pt}
  \small\begin{algorithmic}
    \STATE additional\_mass $\gets$ new\_mass - item.mass
    \STATE item.mass $\gets$ new\_mass
    \STATE item.subtree\_mass += additional\_mass
    \WHILE {item != heap.root \&\& parent(item).mass < item.mass}
    \STATE parent(item).subtree\_mass += additional\_mass
    \STATE heap\_exchange(heap, item, parent(item))
    \STATE item $\gets$ parent(item);
    \ENDWHILE
    \WHILE {item != heap.root}
    \STATE parent(item).subtree\_mass += additional\_mass
    \STATE item $\gets$ parent(item);
    \ENDWHILE
  \end{algorithmic}
  \caption{The Increment operation of the heap-based tree structure.
    The constant-time operation heap\_exchange is demonstrated in
    Figure \ref{heapexchange}.  The two while loops collectively
    over an item's $O(\text{log}|V|)$ ancestors, so Increment is a $O(\text{log}|V|)$ operation.}
  \label{heapincrement}
\end{figure}

\begin{figure}
  Insert(heap, item):
  \small\begin{algorithmic}
    \STATE heap.add(item)
    \STATE node\_mass $\gets$ item.node\_mass
    \WHILE {item has a parent}
    \STATE item $\gets$ parent(item)
    \STATE item.subtree\_mass += node\_mass
    \ENDWHILE
  \end{algorithmic}
  \caption{The Insert operation of the heap-based tree structure.  Note that each item has $O(\text{log} |V|)$ ancestors, so Insert is a $O(\text{log} |V|)$ operation.}
  \vspace*{-15pt}
  \label{heapinsert}
\end{figure}

We use Price's model as an illustrative example.  Recall that, in Price's model, a new node is introduced at each time-step, and an edge from the new node to an existing node is added preferentially.  We first identify an existing node via Sample.  We then create a new node and add an edge from the new node to the existing node.  Increment is called on the existing node to reflect the change in preference mass due to the new incoming edge.  Finally, the new node is added to the index via Insert.  Sample, Increment and Insert are $O(\text{log}|V|)$ operations, which implies that a single iteration is $O(\text{log}|V|)$, and that a simulation with $|V|$ iterations is $O(|V| \text{log} |V|)$.  Generating a network with $|V|$ nodes takes $\Theta(|V|)$ iterations, so $T(G) = O(|V| \text{log} |V|)$.

%


The treap-based implementations were designed to make more efficient use of space.  The heap was implemented via a dynamic array, which provides amortized constant-time insertion at the cost of some wasted space.  We sought to avoid this wastage by instead using binary treaps, which are extensions of binary trees that maintain heap invariants over random priorities to ensure balance in expectation \cite{Aragon:1989gp}.  We find that the treap-based implementation consistently underperformed the heap-based implementation, so we devote less space to its description.

\begin{figure}
  \centering
  \hspace*{-15pt}
  \includegraphics[scale=0.45]{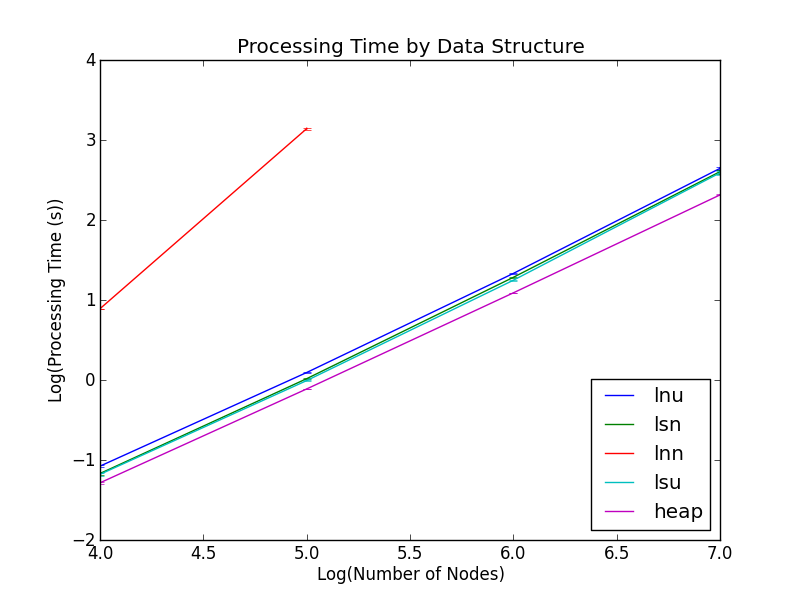}
  \caption{The empirical run time of the simulator using different index types.  The acronyms in the legend indicate the preference function type, the sort key, and the definition of priority for different variants of the treap structure.
    Error bars, barely visible, indicate the 95\% confidence interval.
    }
  \vspace*{-10pt}
  \label{runtime}
\end{figure}

Figure \ref{runtime} shows the empirical run time of each of these structures as a function of generated network size.  All networks were generated from Krapivsky's model.  We find that the binary heap consistently took significantly less time than the treap-based methods to generate networks of several different sizes.



\section{Applications}
\label{sec:applications}
\begin{figure}
  \centering
  \includegraphics[scale=0.32]{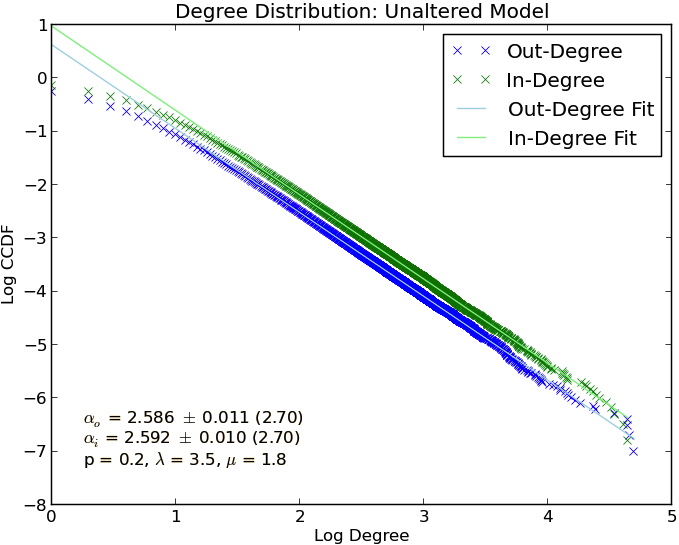}
  \includegraphics[scale=0.32]{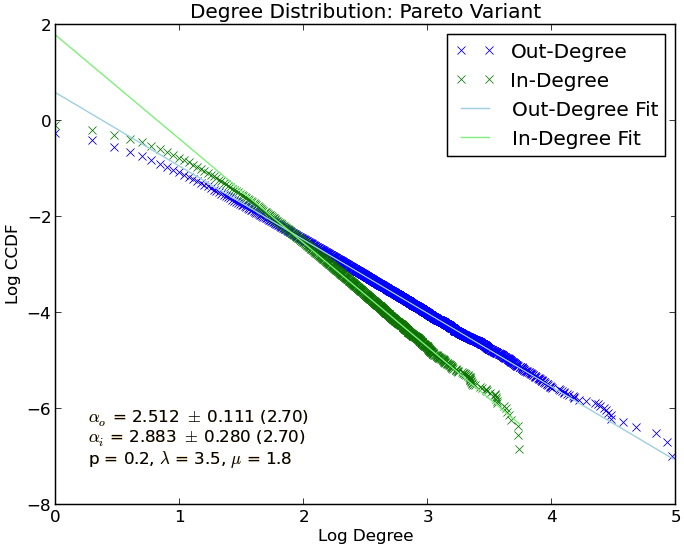}
  \includegraphics[scale=0.32]{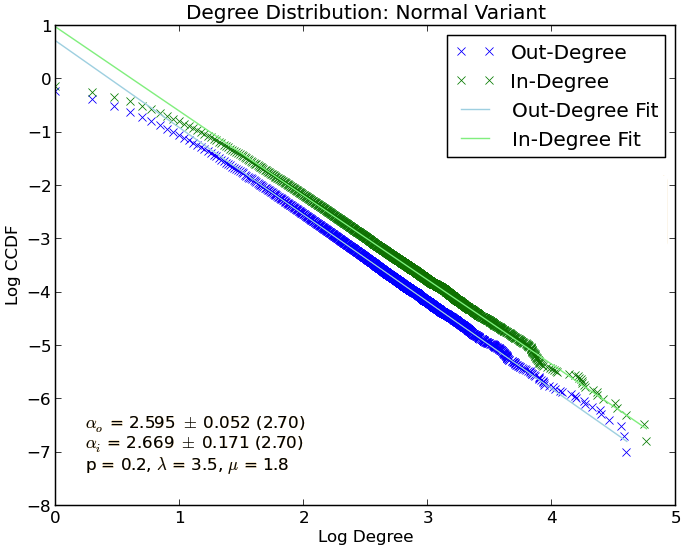}
  \caption{A comparison of the generated and theoretical marginal distributions of in- and out-degree under Krapivsky's model.  The model was parameterized with $\lambda=3.5$ and $\mu=1.8$ and 10 networks with $10^7$ nodes were generated.  We plot the degree distribution of a single example network.  Three variants are investigated: The unaltered model (top), a model with Pareto-distributed fitnesses (center), and a model with normally-distributed fitnesses (bottom).  $\alpha_o = \mu_{o} \pm 2 \sigma_{SE} (\alpha_o^*)$ specifies the inferred exponent of the marginal out-degree distribution, where $\mu_{o}$ is the observed mean of the exponent, $\sigma_{SE}$ is the standard error, and $\alpha_{o}^*$ is the predicted exponent for the unaltered model.  The same form holds for $\alpha_i$.}
  \label{krapivall}
\end{figure}

We validate our generation model by generating sets of networks from the Krapivsky model and comparing the \\marginal degree distributions inferred from the generated networks with the asymptotic value predicted by the model.  We then use the generator to explore some interesting questions.  Specifically, we analyze the effect of changing the fitnesses of the Krapivsky model from a constant value to a random variable with various distributions.  We also analyze the robustness of Krapivsky's model to superlinear preference functions.

\subsection{Validating the Network Generator}


We validate our framework by comparing the inferred exponents of the marginal distributions of generated networks with the known (theoretical, asymptotic) values for the exponents.  We generated 10 networks with $10^7$ nodes each.  Figure \ref{krapivall} shows a plot of the base-10 logarithm of both degree an complementary cumulative distribution.  The exponents of the marginal degree distributions were inferred via linear regression.  We find, as expected, that they both exhibit power-law behavior (evident in the linearity) and that the inferred exponents of the distributions are in relatively good agreement with asymptotic theoretical values.  Note that, while networks with $10^7$ nodes are very large, they are still finite; we believe that this accounts for the small discrepancy between the inferred exponents and the theoretic values.


\subsection{Exploring Extensions to Krapivsky's Model}

\subsubsection{Pareto Fitness}

We use our network generator to investigate the effects of altering the Krapivsky model.  Specifically, we generated networks from a variant where the fitnesses assigned to each node were sampled from a Pareto distribution, rather than assigning the same constant value to each node.  Results can be seen in Figure \ref{krapivall}.  The distribution of in-degree fitness is $\frac{\lambda d_m^\lambda}{d^{\lambda + 1}}$ and has expected value $\frac{\lambda d_m}{\lambda - 1}$.  The parameter $d_m$ is set to $(\lambda - 1)$ so that the expected value of the distribution simplifies to $\lambda$.  The same form was used for the out-degree fitness.  Note that this variant still exhibits scale-free behavior, that the inferred exponents are in better agreement with the predicted values than the exponents inferred from the simulation of the unaltered model, and that the variance of the inferred exponents is higher.


\subsubsection{Normal Fitness}
We also simulated a variant of the Krapivsky model where fitnesses were sampled from a truncated normal distribution.  Results can be seen in Figure \ref{krapivall}.  In-degree fitnesses were sampled from N($\lambda$, ($\lambda/4)^2)$ and out-degree fitnesses from N($\mu$, $(\mu/4)^2$).  The variances were chosen such that the probability of sampling a negative fitness is very small (less than $10^{-4}$); the distributions were truncated so that any negative samples were replaced with zero.  Note that scale-free behavior is still observed and that the inferred exponents of the marginal distributions of in and out-degree are in very close agreement with the simulation of the original model.



\subsubsection{Robustness to Superlinear Preference Functions}

\begin{figure}
  \centering
  \hspace*{-15pt}
  \includegraphics[scale=0.45]{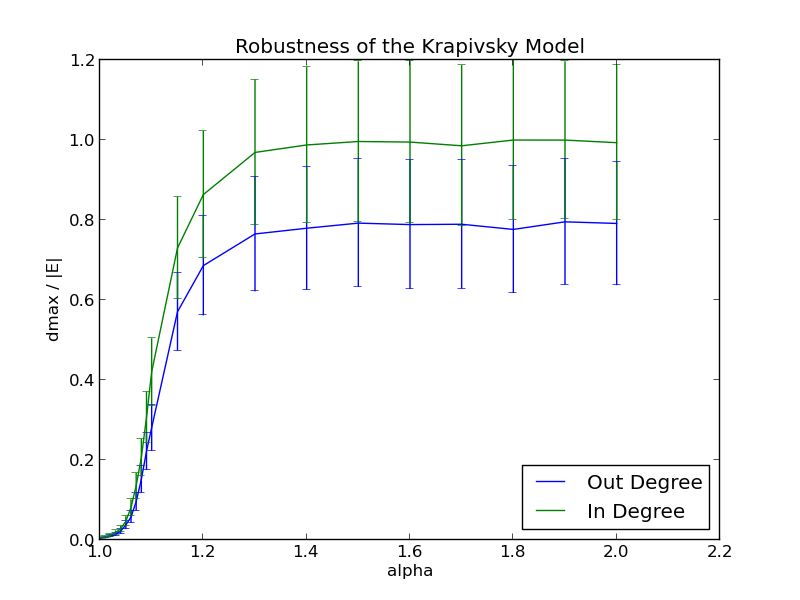}
  \caption{The robustness of the Krapivsky model to superlinear preference functions.  $\alpha$ indicates the exponent of a preference function of the form $f(d) = d^\alpha + c$.  $d_{max}$ is the maximum degree the generated network; $\frac{d_{max}}{|E|}$ gives the proportion of edges that involve the maximum degree node.  $\frac{d_{max}}{|E|} = 1$ indicates a star formation.  Note that there is a phase transition from scale-free to star-structured networks between $\alpha=1.0$ and $\alpha=1.2$.
    100 networks with $10^6$ nodes each were generated for each value of $\alpha$.  Error bars indicate the 95\% confidence interval.
    }
  \vspace*{-10pt}
  \label{robustness}
\end{figure}

Super-linear preference functions increase the strength of the ``rich-get-richer'' effect.  This can lead to situations where one node quickly overtakes all others and is thus a component of most of the edges in the network.  In the extreme case, a star will form; all edges will be connected to the outlier node.  We investigate the robustness of Krapivsky's model to super-linear preference functions by plotting the ratio $\frac{d_{max}}{|E|}$, were $d_{max}$ is the maximum degree, as a function of the preference function exponent $\alpha$; see Figure \ref{robustness}.  $\frac{d_{max}}{|E|}$ will approach 1 as the network approaches a star formation.

There is an interesting side effect to the transition from scale-free to star-structured networks.  As the network becomes more star-like, the probability of selecting the most probable node tends to increase.  The most probable node always sits at the top of the heap, so it can be accessed in constant time.  So, the closer a network's structure is to a star formation, the larger the probability that an iteration of a PA algorithm will be constant time.  For a star structured network in the limit, every iteration will be constant time and the generation of a network with $|V|$ nodes will be $\Theta(|V|)$.  This behavior is apparent for finite $|V|$; we have observed that the runtime of the generator tends to decrease as $\alpha$ increases.

%

\section{Related Work}
\label{sec:relatedwork}
This work is concerned with the problem of efficiently generating networks from PA models.  Some examples of PA models include the models of Price (directed networks with scale-free in-degrees) \cite{Price:1976vn}, Barabasi and Albert (undirected networks with scale-free degrees) \cite{Barabasi:1999uu}, Krapivsky et al. (directed networks with non-independent in and out-degrees which exhibit marginally scale-free behavior) \cite{Krapivsky:2001vj}, and Capocci et al. (like Krapivsky's model, but with reciprocation) \cite{Capocci:2006ju}.

There has been some prior work in efficiently generating networks from PA models.  Tonelli et al. \cite{Tonelli:2010uv} provide a method for computing an iteration of the linear Yule-Simon cumulative advantage process in constant time.  This method can naturally be extended to network generation through linear PA.  However, the extension to nonlinear PA (not shown due to space constraints) is very inefficient in both time and space.  Ren and Li \cite{Ren:2009up} describe the simulation of a particular linear PA model, RX, but do not address the general problem of simulating networks from models with general preference functions.  Hruz et al. \cite{Hruz:2010gp} and D'Angelo and Ferreti \cite{DAngelo:gx} provide methods for parallelizing the simulation of linear PA, but do not treat the nonlinear case.  To the best of our knowledge, our work is the first to address the efficient generation from PA models under possibly nonlinear preference functions.

\section{Conclusion}
\label{sec:conclusion}
We provide an efficient framework for simulating preferential
attachment under general preference functions which scales to millions
of nodes.  We validate this framework empirically and show
applications in the generation and comparison of large networks.

\section{Future Work}
\label{sec:futurework}

We have shown that, for nonlinear preferential attachment, the complexity of generating a network with $|V|$ nodes is both $\Omega(|V|)$ and $O(|V| \text{log} |V|)$.  Future work could provide asymptotically tighter bounds.

\section{Acknowledgments}
This work was supported by MURI ARO grant 66220-9902 and NSF grant CNS-1065133.

\bibliography{simplex2014}
\bibliographystyle{abbrv}
\end{document}